\documentstyle[aps,epsf]{revtex}

\newcommand{\balpha}{{\mbox{\boldmath$\alpha$}}}

\newcommand{\be}{\begin{eqnarray}}
\newcommand{\ee}{\end{eqnarray}}
\newcommand{\la}{\langle}
\newcommand{\ra}{\rangle}

\newcommand{\bfx}{{\bf x}}

\newcommand{\bfr}{{\bf r}}

\newcommand{\bfA}{{\bf A}}

\newcommand{\bfn}{{\bf n}}

\newcommand{\bfp}{{\bf p}}
\newcommand{\veps}{\varepsilon}

\begin{document}
\title
{Recoil correction to the bound-electron $\bf g$ factor in H-like atoms to all orders
in $\balpha \bf Z$ }
\author{V. M. Shabaev and V. A. Yerokhin}

\address
{Department of Physics, St.Petersburg State University, Oulianovskaya 1,
Petrodvorets, St.Petersburg 198504, Russia}
\maketitle
\begin{abstract}
The nuclear recoil correction to the bound-electron $g$ factor in H-like atoms is
calculated to first order in $m/M$ and to all orders in $\alpha Z$. The calculation is
performed in the range $Z=1-100$. A large contribution of terms of order $(\alpha
Z)^5$ and higher is found. Even for hydrogen, the higher-order correction exceeds 
the $(\alpha Z)^4$ term, while for uranium it is above the leading $(\alpha Z)^2$
correction.
\end{abstract}

Recent progress in high-precision measurements of the bound-electron $g$ factor for
H-like carbon \cite{gsi1,gsi2} and the related theoretical investigations
\cite{blu97,per97,bei00a,bei00b,cza01,kar00b,sha01,yel01,gsi3} provide a new
independent determination of the electron mass. The accuracy of this
determination presented in
\cite{gsi3} is three times better than that of the accepted value for
the electron mass \cite{moh00}. This result can be improved if the theoretical and
experimental uncertainties for the $g$ factor are reduced. From the experimental side,
an improvement of the accuracy by an order of magnitude is anticipated in the near
future, as well as an extension of the measurements to higher-$Z$ systems
\cite{wer01}. Investigations of the bound-electron $g$ factor in high-$Z$ systems are
of particular importance since they can provide a new determination of the fine
structure constant \cite{kar00b,wer01},
 nuclear magnetic moments \cite{wer01}, and nuclear
charge radii. They would also create a good possibility for testing the magnetic
sector of QED in a strong Coulomb field. At present, the theoretical uncertainty of
the bound-electron $g$ factor in H-like ions is mainly determined by four factors: a
numerical error in evaluation of the QED correction of first order in $\alpha$, an
error caused by employing the $\alpha Z$ expansion for the QED correction of second
order in $\alpha$, an error resulting from the $\alpha Z$ expansion of the nuclear
recoil correction, and, for very heavy ions, an error due to the finite nuclear size
correction. The main goal of this Letter is to evaluate the nuclear recoil correction
to the $1s$ $g$ factor to all orders in $\alpha Z$ and, therefore, to eliminate one of
the main sources of the uncertainty for the corresponding theoretical predictions.

As is known \cite{lam52}, in the nonrelativistic limit the recoil correction to the
$1s$ $g$ factor vanishes. The leading relativistic recoil correction is of order
$(\alpha Z)^2m/M$ and was evaluated in \cite{fau70a,gro70b} (see also \cite{fau00} and
references therein). General formulas for the nuclear recoil effect valid to all
orders in $\alpha Z$ were derived recently in \cite{sha01}. These results were
confirmed by Yelkhovsky \cite{yel01} by employing a different method. In addition,
Yelkhovsky presented  some arguments for the assertion that the recoil correction up
to order $(\alpha Z)^4 m/M$ is completely
 defined by the so-called lower-order term
(which was evaluated analytically in \cite{sha01} and re-derived in \cite{yel01}). As
a result, in \cite{yel01} the total theoretical uncertainty for the $g$ factor in
C$^{5+}$ was reduced to the level of $1.2\times 10^{-9}$. This leads to improving the
precision of the electron-mass determination by factor of two. In the present Letter,
we numerically evaluate the higher-order contribution to the recoil correction to all
orders in $\alpha Z$. Our results confirm the statement of \cite{yel01} that the
expansion of this term starts with $(\alpha Z)^5$. However, 
we find that the $(\alpha Z)^5$ behavior of the higher-order term is
a result of a cancellation of terms of order $(\alpha Z)^3$ and
$(\alpha Z)^4$ (see a discussion below), while the argumentation of
\cite{yel01} does not contain any indication of the appearance
of such terms.
We also observe that for all H-like atoms in the range $Z=1-100$ 
the higher-order term exceeds the
$(\alpha Z)^4m/M$ contribution. In particular, for the case
 of carbon this
term is about five times larger than the $(\alpha Z)^4 m/M$ term and by factor
of ten exceeds its estimation given in \cite{yel01}.

We consider a H-like atom with a spinless nucleus that is put into the classical
homogeneous magnetic field, ${\bf A}_{\rm cl}(\bfr)=[{\bf {\cal H}}\times {\bfr}]/2$.
For simplicity, we assume that $ {\bf {\cal H}}$ is directed along the $z$ axis. The
energy shift of a state $a$ to first order in ${\cal H}$ and to first order in $m/M$
is conveniently written as the sum of the lower-order and the higher-order term
\cite{sha01,yel01}, $\Delta E=\Delta E_{\rm L}+\Delta E_{\rm H}$, where
($\hbar=c=1\,,\;\;e<0$) \be \label{eq2} \Delta E_{\rm L}&=&\frac{1}{M} \la \delta a|
\Bigr[ \bfp^2-\frac{\alpha Z}{r}(\balpha\cdot\bfp +(\balpha\cdot
\bfn)(\bfn\cdot\bfp))\Bigr]|a\ra\nonumber\\ &&+\frac{e}{2}\frac{{\cal H}}{M}\la
a|\Bigl([\bfr\times\bfp]_z -\frac{\alpha Z}{2r}[\bfr\times\balpha]_z\Bigr)|a\ra\,, \\
\Delta E_{\rm H}&=&\frac{i}{2\pi M} \int_{-\infty}^{\infty} d\omega\;\Bigl\{ \la
\delta a|\Bigl(D^k(\omega)-\frac{[p^k,V]}{\omega+i0}\Bigr)
G(\omega+\veps_a)\Bigl(D^k(\omega)+\frac{[p^k,V]}{\omega+i0}\Bigr)|a\ra \nonumber\\
&&+\la a|\Bigl(D^k(\omega)-\frac{[p^k,V]}{\omega+i0}\Bigr)
G(\omega+\veps_a)\Bigl(D^k(\omega)+\frac{[p^k,V]}{\omega+i0}\Bigr) |\delta
a\ra\nonumber\\ &&+ \la a|\Bigl(D^k(\omega)-\frac{[p^k,V]}{\omega+i0}\Bigr)
G(\omega+\veps_a)(\delta V-\delta \veps_a)\nonumber\\ &&\times G(\omega+\veps_a)
\Bigl(D^k(\omega)+\frac{[p^k,V]}{\omega+i0}\Bigr)|a\ra \Bigr\}\,. \label{eq3}
\ee
Here, $p^k=-i\nabla^k$, $\bfn=\bfr/r$,
 $V(r)=-\alpha Z/r$ is the Coulomb
potential of the nucleus,
 $\delta V(\bfx)=-e\balpha \cdot\bfA_{\rm cl}(\bfx)$,
$D^k(\omega)=-4\pi\alpha Z\alpha^l D^{lk}(\omega)$,
\be \label{eq4}
D^{il}(\omega,{\bf r})=-\frac{1}{4\pi}\Bigl\{\frac
{\exp{(i|\omega|r)}}{r}\delta_{il}+\nabla^{i}\nabla^{l}
\frac{(\exp{(i|\omega|r)}
-1)}{\omega^{2}r}\Bigr\}\,
\ee
is the transverse part of the photon propagator in the Coulomb gauge,
$G(\omega)=\sum_n|n\ra \la n|[\omega-\veps_n(1-i0)]^{-1}$ is the Dirac-Coulomb Green
function, $\delta \veps_a=\la a|\delta V|a\ra$, $|\delta a\ra=\sum_n^{\veps_n\ne
\veps_a}|n\ra\la n|\delta V|a\ra (\veps_a-\veps_n)^{-1}$, and $\balpha$ is a vector
incorporating the Dirac matrices. In equation (\ref{eq3}), the summation over the
repeated indices ($k=1,2,3$), which enumerate components of three-dimensional vectors,
is implicit. The recoil correction to the bound-electron $g$ factor is defined as
$\Delta g=\Delta E/(\mu_0{\cal H}m_j)$, where $\mu_0=|e|/(2m)$ is the Bohr magneton
and $m_j$ is the angular momentum projection of the state under consideration.

For the $1s$ state, the analytical evaluation of the
lower-order term yields \cite{sha01,yel01}
\be \label{eq10}
\Delta g_{\rm L}=\frac{m}{M}(\alpha Z)^2-\frac{m}{M}
\frac{(\alpha Z)^4}{3(1+\sqrt{1-(\alpha Z)^2})^2}\,.
\ee
The first term in the right-hand side of this equation reproduces the result of
\cite{fau70a,gro70b}, while the second term contributes to order $(\alpha Z)^4$ and
higher.

The higher-order term, defined by equation (\ref{eq3}),
is represented by the sum of the
 Coulomb, the one-transverse-photon, and
the two-transverse-photon contribution,
 $\Delta E_{\rm H}=\Delta E_{\rm H}^{\rm Coul} +\Delta
E_{\rm H}^{\rm tr1}+\Delta E_{\rm H}^{\rm tr2}$. For the $1s$ state, we transform them
to the form appropriate for the numerical evaluation,
\be
\label{eq12} \Delta E_{\rm H}^{\rm Coul}&=&\frac{2}{M}\sum_n^{(\veps_n<0)}
\frac{1}{\Delta_{an}^2}
\delta a|[p^k,V]|n\ra\la n|[p^k,V]|a\ra
\nonumber\\
&&+\frac{2}{M}\sum_n^{(\veps_n<0)} \frac{1}{\Delta_{an}^3}
\la a|[p^k,V]|n\ra \la n|(\delta V-\delta
\veps_a)|n \ra \la n|[p^k,V]|a\ra\nonumber\\
&&+\frac{1}{M}\sum_{n_1,n_2}^{(\veps_{n_1}\ne\veps_{n_2})} \la a|[p^k,V]|n_1\ra \la
n_1|\delta V|n_2 \ra \la n_2|[p^k,V]|a\ra
\frac{1}{\Delta_{n_1n_2}}
\Bigl[\frac{\theta(-\veps_{n_1})}{\Delta_{an_1}^2}
-\frac{\theta(-\veps_{n_2})}{\Delta_{an_2}^2}\Bigr]\,,
\\
\Delta E_{\rm H}^{\rm tr1}&=&-\frac{1}{M}\sum_n^{(\veps_n\ne \veps_a)}
\frac{1}{\Delta_{an}}
[\la \delta a|[p^k,V]|n\ra\la n|D^k(0)|a\ra
+\la  a|[p^k,V]|n\ra\la n|D^k(0)|\delta a\ra] \nonumber\\
&&-\frac{2}{\pi M}\int_0^{\infty} dy\;\sum_n^{(\veps_n\ne \veps_a)}
\frac{1}{y^2+\Delta_{an}^2}
[\la \delta a|[p^k,V]|n\ra\la n|S^k(y)|a\ra
+\la  a|[p^k,V]|n\ra\la n|S^k(y)|\delta a\ra] \nonumber\\
&&-\frac{1}{M}\sum_{n_1}^{(\veps_{n_1}\ne \veps_a)}
\sum_{n_2}^{(\veps_{n_2}\ne \veps_a)}
\frac{1}{\Delta_{an_1}\Delta_{an_2}}
\la a |[p^k,V]|n_1\ra \la n_1|(\delta V-\delta \veps_a)|n_2 \ra
\la n_2|D^k(0)|a\ra \nonumber \\
&&-\frac{2}{\pi M}\int_0^{\infty} dy\;
\sum_{n_1}^{(\veps_{n_1}\ne \veps_a)}
\sum_{n_2}^{(\veps_{n_2}\ne \veps_a)}
\frac{\Delta_{an_1}+\Delta_{an_2}}
{(y^2+\Delta_{an_1}^2)(y^2+\Delta_{an_2}^2)}
\nonumber \\
&& \times
\la a |[p^k,V]|n_1\ra \la n_1|(\delta V-\delta \veps_a)|n_2 \ra
\la n_2|S^k(y)|a\ra\,,
\label{eq13}
\\
\Delta E_{\rm H}^{\rm tr2}&=&-\frac{2}
{\pi M}\int_0^{\infty} dy\;\sum_n^{(\veps_n\ne \veps_a)}
\frac{\Delta_{an}}{y^2+\Delta_{an}^2}
\la \delta a|S^k(y)|n\ra\la n|S^k(y)|a\ra \nonumber \\
&&-\frac{1}{\pi M}\int_0^{\infty} dy\;
\sum_{n_1}^{(\veps_{n_1}\ne \veps_a)}
\sum_{n_2}^{(\veps_{n_2}\ne \veps_a)}
\frac{\Delta_{an_1}\Delta_{an_2}-y^2}
{[y^2+\Delta_{an_1}^2][y^2+\Delta_{an_2}^2]}
\nonumber \\
&& \times
\la a |S^k(y)|n_1\ra \la n_1|(\delta V-\delta \veps_a)|n_2 \ra
\la n_2|S^k(y)|a\ra\,,
\label{eq14}
\ee
where ${\bf S}(y)=\alpha Z \balpha\exp{(-yr)}/r +i\alpha Z[H,\phi(yr){\bf n}]$,
$H=\balpha \cdot \bfp+\beta m +V(r)$ is the Dirac-Coulomb Hamiltonian, $\Delta_{ij} =
\veps_i-\veps_j$, and $\phi(yr)=[\exp{(-yr)}(1+yr)-1]/(yr)^2$. The wave-function
correction $|\delta a \ra$ can be found analytically by the method of generalized
virial relations for the Dirac equation \cite{sha91}. The explicit form for the
component of $|\delta a\ra$, that has the same angular quantum numbers as the
reference state $|a\ra$, is presented in \cite{sha01} (only this component contributes
to the effect under consideration).

The numerical evaluation of the expressions (\ref{eq12})-(\ref{eq14}) was carried out
similarly to our previous calculations of the nuclear recoil correction to the Lamb
shift \cite{art95,sha98b,sha98c}. After integration over angles, the finite basis set
method was used to evaluate infinite summations over the electron spectrum. Basis
functions were constructed from B-splines by employing the procedure proposed in
\cite{joh88}. The integration over $y$ was carried out numerically for equation
(\ref{eq14}) and both numerically and analytically for equation (\ref{eq13}). We
mention large numerical cancellations arising in (\ref{eq13}) for very small $Z$ if
the $y$ integration is performed analytically. In this case, numerical integration
turns out to be preferable. The number of B-splines used in actual calculations was
varied from 70 to 110. The estimated uncertainty corresponds to the dependence of the
results on grid parameters and the number of splines and integration points.

The correction to the $1s$ $g$ factor induced by the higher-order term
$\Delta E_{\rm H}$ is
expressed in terms of the function $P(\alpha Z)$,
\be
\label{eq16} \Delta g_{\rm H}=\frac{m}{M}(\alpha Z)^5 P(\alpha Z)\,.
\ee
The corresponding numerical results are presented in Table I. It is noteworthy that
the one-transverse-photon and the two-transverse-photon contribution separately are of
the order $(\alpha Z)^4$ for small $Z$, while their sum exhibits the $(\alpha Z)^5$
behavior. This fact is clearly demonstrated in Fig. 1, where the numerical results for
the ratio $\Delta g_{\rm H}/[(m/M)(\alpha Z)^4]$ are plotted. We also note that the
one-transverse-photon contribution contains terms of order $(\alpha Z)^3$ which are
cancelled when added together. Namely, the part corresponding to the perturbation of
the reference state $a$ and the part corresponding to the perturbation of the electron
propagator exhibit the $(\alpha Z)^3$ behavior, when taken separately. 
We note that, in contrast to our results, 
the argumentation of \cite{yel01},
where the same gauge is considered, does not indicate
the appearance of terms of order lower than $(\alpha Z)^5$.
 For this reason, we can not consider
the argumentation of Yelkhovsky, in the form it is given in \cite{yel01},
as complete.
 Fitting our numerical results for small $Z$ to
the form $P(\alpha Z)=C_{51}\log{(\alpha Z)}+C_{50} +\alpha Z(\cdots)$ yields
$C_{51}=-5.3\pm 0.5$ and $C_{50}=-6.5\pm 1.0$. The uncertainties of the coefficients
were estimated analyzing the dependence of the results on the number of parameters in
the fit and the number of fitting points.

In Table 2 we present the ratios $\Delta g_{\rm H}/\Delta g_{0}$ and $(\Delta g_{\rm
L}-\Delta g_{0})/\Delta g_{0}$, where $\Delta g_{0}=(\alpha Z)^2 m/M$ is the
lowest-order correction derived in \cite{fau70a,gro70b}. As can be seen from the
table, the $\Delta g_{\rm H}$ term exceeds $(\Delta g_{\rm L}-\Delta g_{0})$ for all
$Z$ in the range $Z=1-100$. In the case of carbon, $\Delta g_{\rm H}$ amounts to
$7.7\times 10^{-11}$, which is ten times larger than the uncertainty ascribed to this
term in \cite{yel01}. However, since this correction is about ten times smaller than
the current theoretical uncertainty due to the binding QED correction, it does not
affect the electron-mass prediction of \cite{gsi3}. The higher-order recoil correction
is more important for higher-$Z$ systems, since it grows very rapidly when $Z$
increases. In particular, for uranium the higher-order recoil correction is even above
the $(\alpha Z)^2 m/M$ term.

In Table 3 we present the individual contributions to the $1s$ $g$ factor for some
H-like ions in the range $Z=6-92$. An error ascribed to the Dirac point-nucleus value
results from the current uncertainty of the fine structure constant,
$1/\alpha=137.03599976(50)$ \cite{moh00}. The uncertainty of the finite nuclear size
correction was estimated as the difference between the result obtained with the Fermi
model of the nuclear charge distribution and with the homogeneously-charged-sphere
model. The nuclear charge radii were taken from \cite{fri95,zum84}. The one-loop QED
correction was taken from \cite{per97,bei00a}, where it was evaluated numerically to
all orders in $\alpha Z$ . The $\alpha^2$ QED correction includes the existing $\alpha
Z$ expansion terms for the QED correction of second order in $\alpha$
\cite{cza01,kar00b} and the known free-QED terms of higher orders in $\alpha$ (see,
e.g., \cite{bei00b}). Its relative uncertainty was estimated as the ratio of the part
of the one-loop QED correction, that is beyond the $(\alpha Z)^2$ approximation, to
the part, that is within the $(\alpha Z)^2$ approximation. The recoil correction
incorporates the total recoil correction of first order in $m/M$, calculated in this
work, and the known corrections of orders $(m/M)^2$ and $\alpha(m/M)$ \cite{fau00}.
From the table, we conclude that for low $Z$ the theoretical uncertainty is mainly
determined by the numerical error of the one-loop QED correction \cite{bei00a,bei00b},
while for high $Z$ it results from the $\alpha Z$ expansion of the $\alpha^2 $ QED
correction and from the finite nuclear size correction. Calculations of the QED
corrections up to the desirable accuracy seem to be feasible in the near future  if we
consider recent progress in calculations of the corresponding corrections to the Lamb
shift in H-like ions \cite{jen99,yer01}. As to the uncertainty due to the finite
nuclear size effect, one may expect that, like in the case of the hyperfine splitting
\cite{sha01b}, it can be significantly reduced in a specific difference of the
bound-electron $g$ factor in H- and Li-like ions.

We thank  T. Beier, S. Karshenboim, J. Kluge, W. Quint, and A. Yelkhovsky for valuable
discussions and N. Lentsman for improving the language. This work was supported in
part by RFBR (Grant No. 01-02-17248), by the program "Russian Universities - Basic
Research" (project No. 3930), and by GSI.


\begin{table}
\caption{The higher-order recoil correction to the $1s$ $g$ factor,
expressed in terms of the function $P(\alpha Z)$ defined by
equation (8).}
\begin{tabular}{cr@{.}lr@{.}lr@{.}lr@{.}l} \hline
$Z$& \multicolumn{2}{c}{$P_{\rm Coul}$}
                    & \multicolumn{2}{c}{$P_{\rm tr1}$}
                                     & \multicolumn{2}{c}{$P_{\rm tr2}$ }
                                                     & \multicolumn{2}{c}{$P$}
\\
\hline
  1  &   $-$1&11414   & 100&701(2)     &   $-$80&8200(3) &   18&769(2)  \\
  2  &   $-$1&09754   &  53&5278(6)    &   $-$36&98689   &   15&4434(6) \\
  3  &   $-$1&08183   &  37&44949(6)   &   $-$22&80837   &   13&55928(6) \\
  4  &   $-$1&06693   &  29&24592(4)   &   $-$15&91960   &   12&25940(4) \\
  5  &   $-$1&05277   &  24&23027(3)   &   $-$11&90049   &   11&27702(3) \\
  6  &   $-$1&03931   &  20&82713(1)   &   $ -$9&29387   &   10&49395(1) \\
  8  &   $-$1&01429   &  16&47349      &   $ -$6&15902   &    9&30018(1)  \\
 10  &   $-$0&99161   &  13&78331      &   $ -$4&37646   &    8&41524(1)  \\
 20  &   $-$0&90647   &   8&09979      &   $ -$1&22907   &    5&96425(1)  \\
 30  &   $-$0&85834   &   6&08456      &   $ -$0&41612   &    4&81010(1)  \\
 40  &   $-$0&84048(1) &  5&09672(1)   &   $ -$0&08937   &    4&16687(1) \\
 50  &   $-$0&85209(1) &  4&58136(3)   &       0&07843   &    3&80770(3)  \\
 60  &   $-$0&89803(3) &  4&36161(3)   &       0&18668(1)&    3&65025(4) \\
 70  &   $-$0&99173(9) &  4&39148(9)   &       0&27839(1)&    3&6781(1)  \\
 80  &   $-$1&1647(1)  &  4&7153(3)    &       0&38378(3)&    3&9344(3)  \\
 90  &   $-$1&4962(9)  &  5&525(3)     &       0&5459(2) &    4&575(3)  \\
100  &   $-$2&228(9)   &  7&48(3)      &       0&883(2)   &   6&14(3)  \\
\hline
\end{tabular}
\end{table}

\begin{table}
\caption{The recoil corrections to the $1s$ $g$ factor, expressed in terms of the
leading correction, $\Delta g_{0}=(\alpha Z)^2 m/M$. The difference $\Delta g_{\rm
L}-\Delta g_{0}$ is defined by the second term in equation (4) and corresponds to the
deviation of the lower-order term from $(\alpha Z)^2 m/M$. $\Delta g_{\rm H}$ is the
higher-order term.}
\begin{tabular}{c l l } \hline
$Z$ & $(\Delta g_{\rm L}-\Delta g_{0})/\Delta g_{0}
$ &  $\Delta g_{\rm H}/\Delta g_{0}$
\\
\hline
1& -4.437732$\times 10^{-6}$& 7.2935(8)$\times 10^{-6}$ \\
2& -1.775234$\times 10^{-5}$&4.8010(2)$\times 10^{-5}$ \\
6& -1.599074$\times 10^{-4}$&8.80823(1)$\times 10^{-4}$ \\
10&-4.449468$\times 10^{-4}$&3.27011(1)$\times 10^{-3}$ \\
20& -1.794206$\times 10^{-3}$&1.85414(1)$\times 10^{-2}$ \\
30&-4.092524$\times 10^{-3}$&5.04677(1)$\times 10^{-2}$ \\
50& -1.190031$\times 10^{-2}$&0.184956(2) \\
70& -2.514919$\times 10^{-2}$&0.49025(2) \\
90& -4.672903$\times 10^{-2}$&1.296(1) \\
100& -6.261303$\times 10^{-2}$&2.39(1) \\
\hline
\end{tabular}
\end{table}

\begin{table}
\squeezetable
\caption{The individual contributions to the $1s$ bound-electron
$g$ factor in H-like ions.}
\begin{tabular}{lr@{.}lr@{.}lr@{.}lr@{.}lr@{.}lr@{.}l}
                           &  \multicolumn{2}{c}{$^{12}{\rm C}^{5+}$}
                                                 &   \multicolumn{2}{c}{$^{16}{\rm O}^{7+}$}
                                                                           & \multicolumn{2}{c}{$^{32}{\rm S}^{15+}$}
                                                                                                  & \multicolumn{2}{c}{$^{40}{\rm Ar}^{17+}$}
                                                                                                                           & \multicolumn{2}{c}{$^{40}{\rm Ca}^{19+}$} \\
                    \hline
Dirac value (point)        &   1&998 721 354 4   &     1&997 726 003 1
    &   1&990 880 058 3(1)      &
1&988 447 661 3(1)        &   1&985 723 203 8(1)      \\
Fin. nucl. size            &   0&000 000 000 4   &
  0&000 000 001 5     &   0&000 000 038 9    &   0&000 000 070 3      &   0&000 000 113 1(1)   \\
QED, order $(\alpha/\pi)$  &   0&002 323 663 7(9)&     0&002 324 416(1)    &   0&002 330 920(3)   &   0&002 333 636(4)     &   0&002 336 92(1)      \\
QED, order $(\alpha/\pi)^2$&$-$0&000 003 516 2(2)&
 $-$0&000 003 517 1(4)  &$-$0&000 003 523(4)   &$-$0&000 003 525(6)     &$-$0&000 003 528(9)     \\
Recoil                     &   0&000 000 087 6   &     0&000 000 117 0     &   0&000 000 236 0    &   0&000 000 239 8      &   0&000 000 297 1      \\
Total                      &   2&001 041 589 9(9)&    2&000 047 021(1)    &   1&993 208 254(5)   &   1&990 778 082(8)     &   1&988 057 01(2)      \\
\hline \hline
                           & \multicolumn{2}{c}{$^{52}{\rm Cr}^{23+}$}
                                                    & \multicolumn{2}{c}{$^{74}{\rm Ge}^{31+}$}
                                                                             & \multicolumn{2}{c}{$^{132}{\rm Xe}^{53+}$}
                                                                                                      & \multicolumn{2}{c}{$^{208}{\rm Pb}^{81+ }$}
                                                                                                                               &\multicolumn{2}{c}{ $^{238}{\rm U}^{91+}$}  \\
\hline
Dirac value (point)        &   1&979 392 224 9(2)
  &   1&963 137 509 5(3)        &   1&892 114 650(1)     &   1&734 947 026(2)     &   1&654 846 173(3)  \\
Fin. nucl. size            &   0&000 000 272 6(2)   &   0&000 001 231 2(10)
   &   0&000 023 49(3)      &   0&000 453 3(9)       &   0&001 275 0(25)   \\
QED, order $(\alpha/\pi)$  &   0&002 345 02(1)     &   0&002 369 20(1)      &   0&002 505 26(1)      &   0&002 884 38(3)      &   0&003 088 93(3)   \\
QED, order $(\alpha/\pi)^2$&$-$0&000 003 533(16)   &$-$0&000 003 55(4)      &$-$0&000 003 61(19)     &$-$0&000 003 7(6)       &$-$0&000 003 8(9)    \\
Recoil                     &   0&000 000 332 4     &   0&000 000 426 5      &   0&000 000 783 5      &   0&000 001 723        &   0&000 002 491      \\
Total                      &   1&981 734 32(2)     &   1&965 504 82(4)      &   1&894 640 57(19)     &   1&738 282 7(11)      &   1&659 208 9(27)
\end{tabular}
\end{table}

\begin{figure}

\centerline{ \mbox{\epsfxsize=13.6cm \epsffile{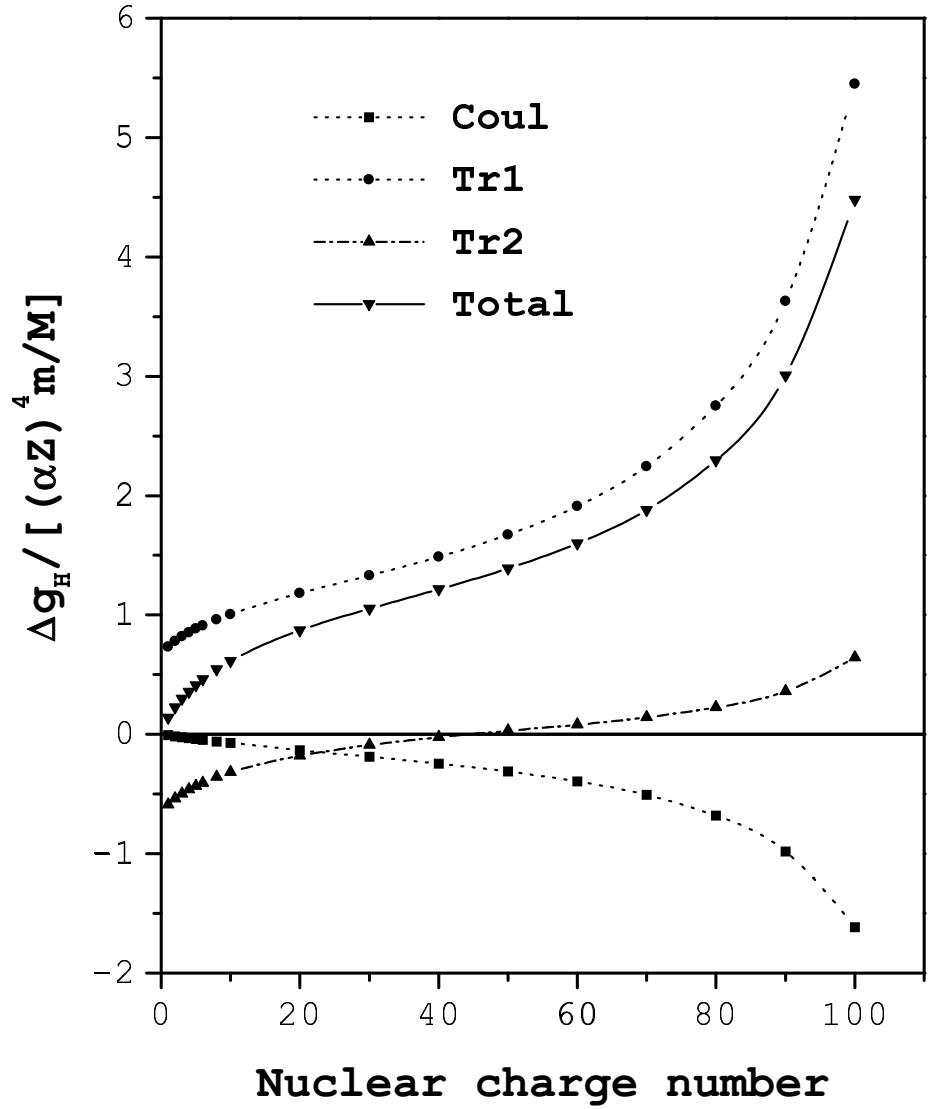}}} \caption{The Coulomb, the
one-transverse-photon, and the two-transverse-photon contribution to the ratio $\Delta
g_{\rm H}/[(\alpha Z)^4m/M]$. }
\end{figure}

\end{document}